\documentclass[twocolumn,showpacs,preprintnumbers,amsmath,amssymb]{revtex4}
\usepackage{graphicx}
\usepackage{dcolumn}
\usepackage{bm}
\begin{document}
\preprint{}
\title{Critical indices of random planar electrical  networks}
\author{J.S.\,Espinoza\,Ortiz$^{\ddagger}$ and Gemunu\,H.\,Gunaratne$^{\star\,\dag}$}
\affiliation{$^{\ddagger}$Departamento de F\'{\i}sica,Universidade Federal de Goi\'as,\,Catal\~ao,\,GO\,75700-000,\,Brazil} 
\affiliation{$^{\star}$Department of Physics,\,University of Houston,\,Houston,\,TX\,77204,\,USA}
\affiliation{$^{\dag}$The Institute of Fundamental Studies,\,Kandy 20000,\,Sri Lanka}

\begin{abstract}
We propose a new method to estimate the critical index $t$ for
strength of networks of random fused conductors. It relies on a
recently introduced expression for their yield strength.
We confirm the results using finite size scaling.
To pursue this commitment, we systematically study  different damage modalities
of conducting networks inducing variations on the behavior of their nonlinear strength reduction.
\end{abstract}
\pacs{87.15.Aa, 87.15.La, 91.60.Ba, 02.60.Cb} 
\maketitle

\section{Introduction}
Random resistor networks are used to model a variety of physical phenomena ranging from the properties of inhomogeneous media\;\cite{redneru,sahimi,chung,ray,kantor}, to metal insulator transitions\;\cite{shante}, dielectric breakdown\;\cite{takayasu,dyre,hansen,hansenin}, the role of percolation in weak 
and strong disorder\;\cite{tsallis,stanley,havlin}, and the strength of trabecular bone\;\cite{fung,faulkner,gunaratne}. Essentially, this study  intends to develop  
a method to establish a relationship between the mean strength of bone and its mass. In fact, large bones consist of an outer compact shaft and an inner porous region, i.e.,
a trabecular architecture whose structure is reminiscent of a complex system of disordered networks. The bone strength depends on many factors, i.e., architectural
characteristic (level of connectivity), perforation, thinning, anisotropy, as well as subarchitectural properties of bone like the mineral content, the density of the diffuse
damage. A significant mechanism for loss of trabecular mass is through the removal of individual struts, due to traumatic events.  Mechanical studies on \textit{ex\,vivo} bone samples have shown that trabecular networks from patients with a broad range of age fracture at a fixed level of strain\;\cite{morgan}, even though the corresponding fracture stresses exhibit large variations. These observations have motivated the induced fracture criteria utilised in our models.

We consider the lattice network of conducting disordered elements to study the essential non-linear and irreversible properties of the electrical breakdown strength. Subjecting the system under extreme perturbation,  its electrical  properties tend to get destabilised and so that failure breakdown occurrences. In fact, these instabilities in the system often nucleates around disorder, which plays a major role in the breakdown properties of the system. The growth of these nucleating centers, in turn, depends on various statistical properties of the disorder, namely the scaling properties of percolation structures, its fractal dimensions, etc. By increasing the percentage number of the network nonconducting elements,  its conductivity decreases, so also does the breaking strength of the material, the fuse current of the network decreases on the average with the increased concentration of random impurities.

We are considering here  the problem associated with failures in disordered system under the influence of an electrical field. In spite of  this framework  is shown simpler than the one related to mechanical failure of fractures, however all these cases of failures present some common features. In particular, certain features near breakdown agree with those of resistor networks close to the percolation threshold\;\cite{CB,duxbury,duxburyin,herrmann}. This article presents a systematic new method to estimate critical indices by quantifying the reduction of current in fused resistor networks. 

The model and the expression which relate the strength reduction to the statistical properties of disorder network system is set up in section II. We focus on fractured conducting networks to examine many possibilities for. Initially, it is supposed that disorder just arises from random percolation. Both cases are considered, the isotropic and the anisotropic conducting networks. Next,  it is also considered that the failure current of the conducting network are not  the same, but it is being uniformly distributed in a range, in addition to random uniform percolation. This issues are presented and quantified in Section III. Finally, Section IV is devoted to conclusions.

\section{The Model}
We take square networks consisting of fused conductors that fail when the potential difference across them reaches a pre-set value; {\it\,i.e.}, the breakdown current
of an element is proportional to its conductance. Typically, failure of an element increases currents on neighboring conductors, enhancing the likelihood of their failure\;\cite{essam,stauffer}. We study the yield point at which the external current initiates the first failure. The peak currents on a network show similar behavior\;\cite{DGM}.

Consider first, a complete square network of size $M\times\,M$, with the top and the bottom edges at potentials $V_{0}$ and $0$
respectively. Assume that each electrical element in the network fails when the potential difference across it reaches a value $V_{b}$.
We can calculate the current $I(0)$ flowing through the network using Kirchhoff's laws. As $V_{0}$ increases, currents through the
conductors, as well as $I(0)$, will increase until the yield point $I(0) = I_{yield}(0)$, where the first failure occurs.

Next consider a network, where we have attempted to remove elements with a probability $p$. Denote the yield current of such
a network by $I_{yield}(p)$. Typically, $I_{yield}(p)$ decreases with increasing $p$, and vanishes as $(p_{0}-p)^{t}$ when $p$
approaches the percolation threshold $p_{0}\,$(=0.5 for isotropic square networks). $t$ is the critical index, with reported values
between $1.1$ and $1.43$ under different scenarios of damage and symmetries of the network \cite{lobbc,kirkpatrickc,kirkpatrickc2,gefenc,benguiguic,halperingc,frankc,halperingc2,sahimib}.

There have been several proposals related to the form of the reduction of strength of a network due to a random removal of element\cite{duxbury,CB}. Here we test  a recently proposed expression for $I_{yield}(p)$\;\cite{ortiz}:
\begin{eqnarray}
\tau(p)\equiv\frac{I_{yield}(p)}{I_{yield}(0)}=\frac{1}{1\,+\,a_{1}\,z\,^{t/2}\,+\,a_{2}\,z\,^{t}}\,,
\end{eqnarray}
where $z=\log(N)/\log(\frac{p_{0}}{p})$. Here, $N(=M^{2})$ is the number of nodes in the original network, $p_{0}$ is the bond
percolation threshold for the class of network considered, and $a_{1}$ and $a_{2}$ are constants that depend on model parameters
as discussed below. Observe that, as $p\rightarrow p_{0}$, the yield strength, $\tau(p)\rightarrow(p_{0}-p)^{t}$. We conjecture
that Equation (1) is valid throughout the range $p\in[0,p_{0})$, we use this conjecture to estimate $\,p_{0}$ and the critical index $t\,$ ; then validate it using finite size scaling method.  

We compute the yield current for a given network as follows: given the conductance $\sigma_{i}$ of all electrical elements in the
network and the potential $V_{o}$ of the top layer of nodes, we use Kirchhoff's laws to determine the currents $i_{k}$ through each element and the potential differences $v_{k}$ across them. We denote the largest of the latter by $V_{max}$. The current $I(p)$
passing through the network is the average of all currents through electrical elements belonging to a fixed horizontal layer. The
yield current is $I_{yield}(p)=I(p)\times\,V_{b}/V_{max}$.

\begin{figure}[t]
\centering \vspace*{0.25cm} \setlength{\abovecaptionskip}{0.2cm}
\includegraphics[width=8cm,height=7.45cm]{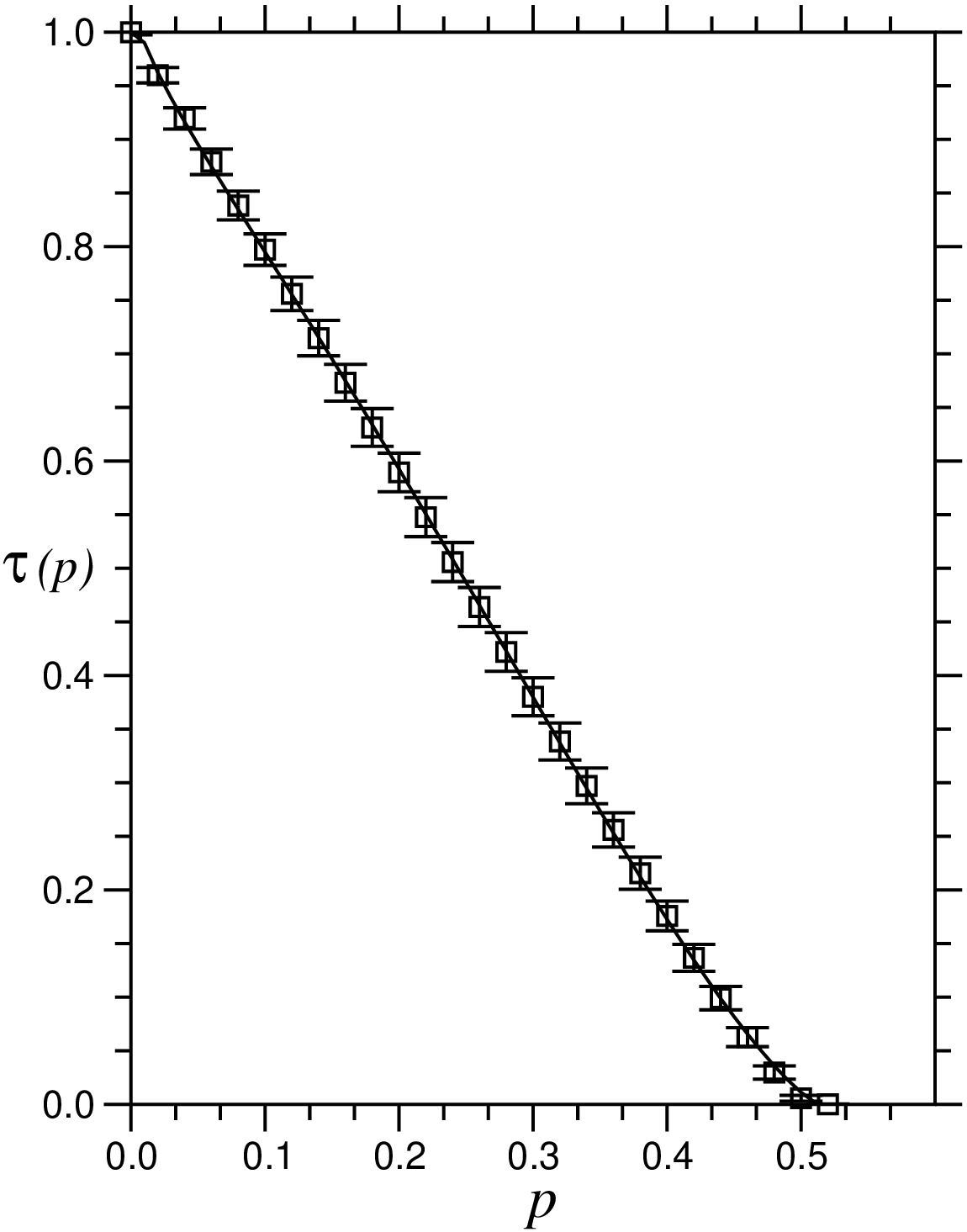}
\caption{\label{fig:epsart1} Strength reduction in a $160^{2}$
network due to a random isotropic removal of a fraction $\,p\,$ of
conductances, averaged over one thousand trials. Numerical results
are shown by boxes along with the statistical error, while those
obtained by fitting to expression (1) are shown as a continuous
line.}
\end{figure}

\begin{figure}[b]
\centering \vspace*{0.2cm} \setlength{\abovecaptionskip}{0.2cm}
\includegraphics[width=7.75cm,height=7.25cm]{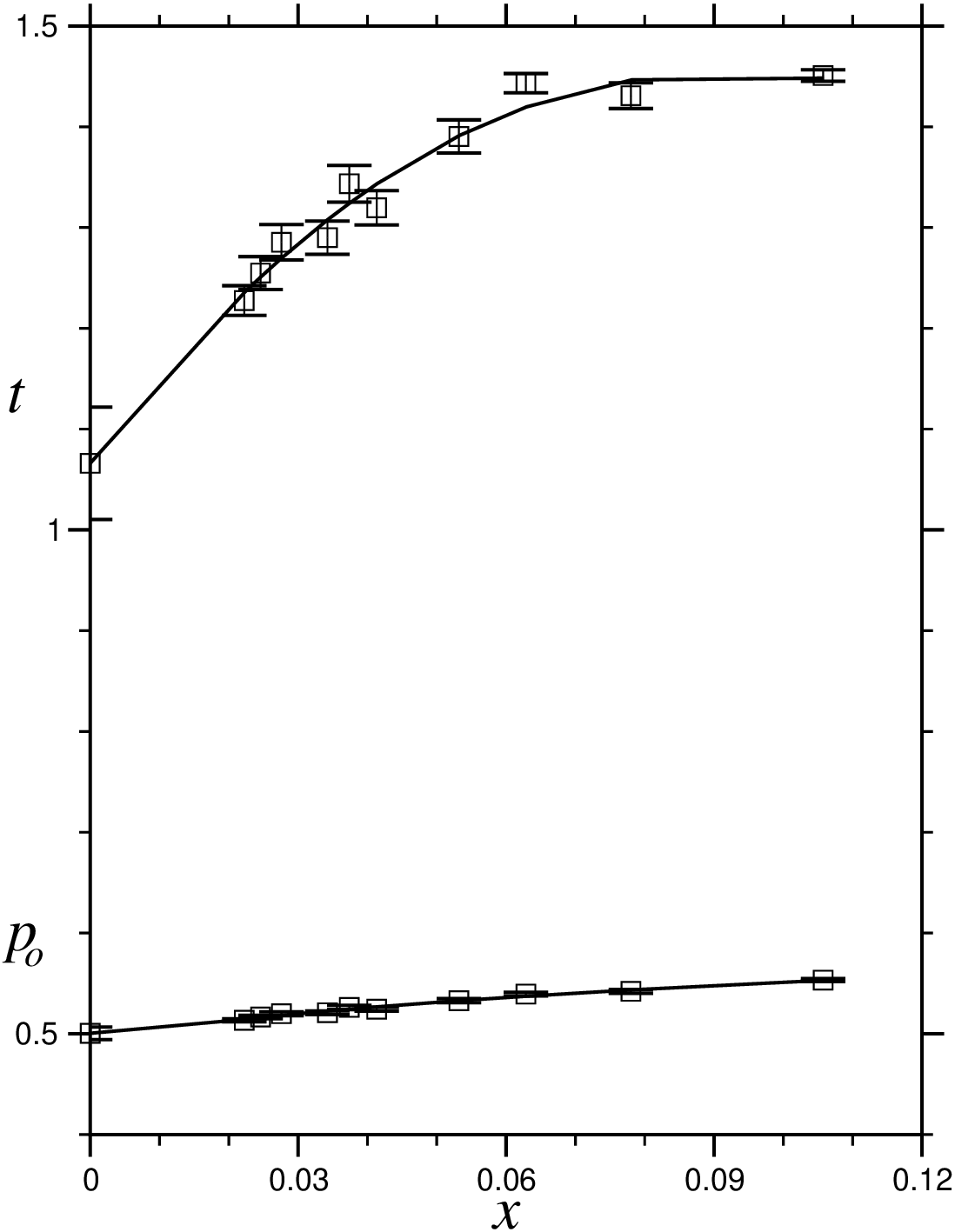}
\caption{\label{fig:epsart2} Variation of parameters $t$ and $p_0$
versus the inverse correlation length $(\,x=M^{-0.75}\,)$, for
networks of size \,$M=20$ to $M=160$. All elements in the initial
networks have conductance of 1 unit. Note that, both critical values for
$t\,(=1.066\pm\,0.056)$ and $p_{0}\,(=0.500\pm\,0.006)$ are getting in the limit when $x\rightarrow\,0$. }
\end{figure}

\section{Results and Discussions}
Our computations used networks with sizes $M=20,\,30,\,40,\,50,\,70,\,80,\,90,\,120,\,140$ and $160$, and sample sizes of $10,000$ for $M=20,\,30,\,40,\,50$, sample size of $2,500$ for $M=70,\,80,\,90$ and finally sample size of $1,000$ for $M=120,\,140,\,160$. Figure~\ref{fig:epsart1} shows the behavior of $\tau(p)$ for the $160\times\,160$ networks, where the error bars show standard errors. Although fluctuations $\delta\tau$ in $\tau(p)$ decrease as $p\rightarrow p_{0}$,
the relative fluctuations ($\delta\tau/\tau$) increase. The solid line shown in Figure~\ref{fig:epsart1} represents the best fit to Equation (1)
with parameters $a_{1},\,a_{2},\,p_{0}$ and $t$ in a $160^{2}$ conducting network. To determine their values, $a_{1}=-0.1043\pm\,0.005,\,a_{2}=0.061\pm\,0.003, \,p_{0}=0.513\pm\,0.002$ and $t=1.228\pm\,0.015$, we used the Lavenberg-Marquadt method to implement the nonlinear fit \cite{recipes}. We must fit $t$ and $p_{0}$ because they depend on the size of the network.

Next, we determine how $p_{0}$ and $t$ change with the network size. We express the parameters as a function of $x=M^{-1/\nu}$ where $\nu$ ($=4/3$ for 2D square networks) is the universal correlation length exponent\;\cite{nijs,yu}; {\it\,i.e}, $x$ is the inverse of the mean size of the largest domain in a network of size $M\times\,M$. Figure~\ref{fig:epsart2} shows the values of $t(x)$ and $p_{0}(x)$, along with the error estimates.

We estimate the value of $t(0)$ corresponding to an infinite
network by first approximating $t(x)$ by a rational function
$f(x)/g(x)$ (where $f(x)$ and $g(x)$ are polynomials of order $3$
and $2$ respectively) and extrapolating to $x=0$. The values of
the extrapolation corresponding to Figure~\ref{fig:epsart2} are
$p_{0}(0)=0.500\pm\,0.006,\,t(0)=1.066\pm\,0.056,\,a_{1}(0)=-0.106\pm\,0.01
5\,$ and $\,a_{2}(0)=0.060\pm\,0.004$. The error estimate includes
both errors at each $M$ and those due to the extrapolation
\cite{recipes}.

\begin{figure}[t]
\centering\hspace*{-0.925cm} \vspace*{0.2cm} \setlength{\abovecaptionskip}{0.2cm}
\includegraphics[width=8.75cm,height=5.75cm]{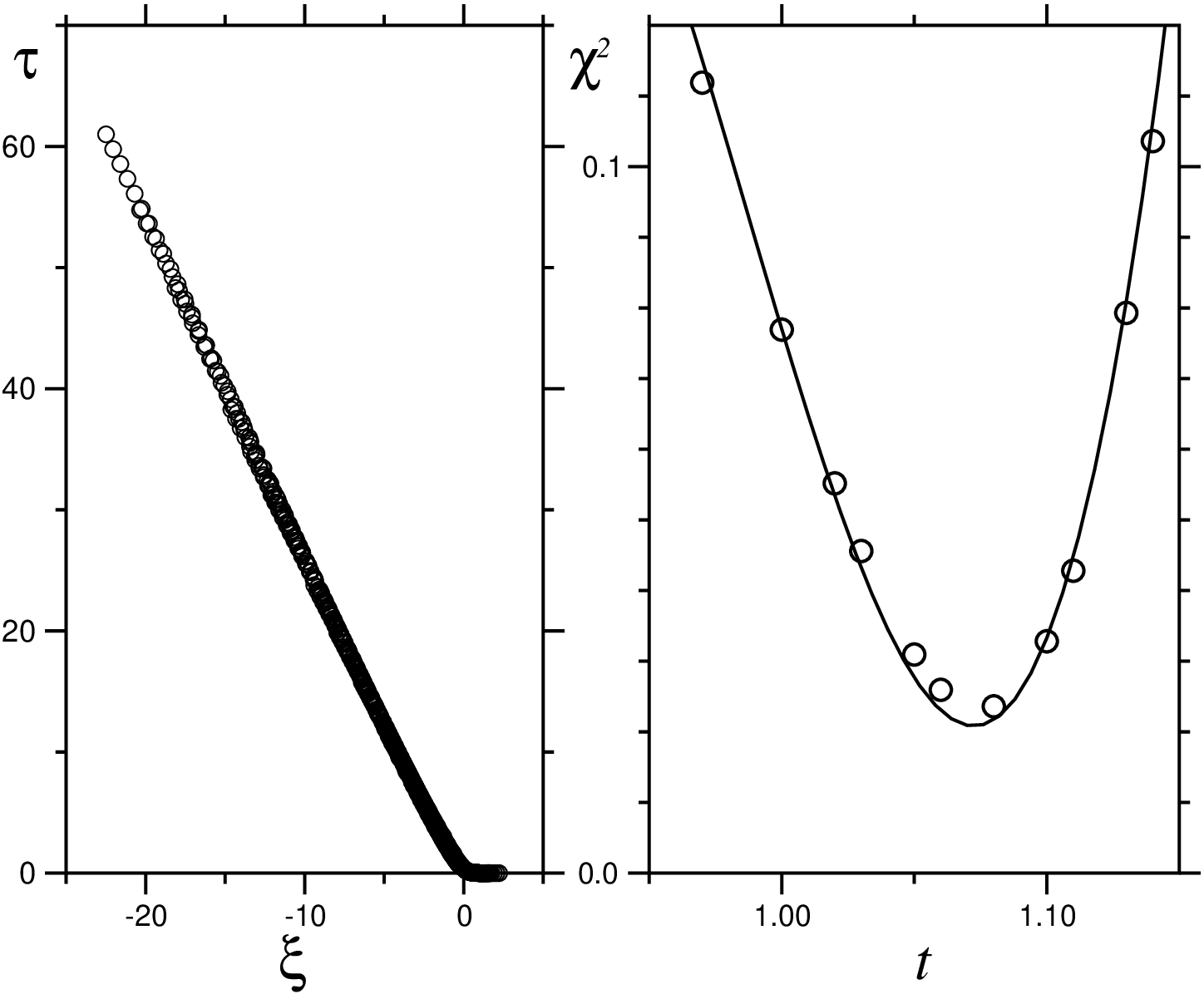}
\caption{\label{fig:epsart3} The left side of the figure shows the scaling
function ($\tau$) behavior as a function of the network
correlation length ($\xi$), and its $\chi^{2}$ error as a function
of $t$ for fixed unit conductances in the network (right side).
The minimum of $\chi^{2}$ is $t$, gives the critical exponent.}
\end{figure}

\begin{figure}[b]
\centering \vspace*{0.2cm} \setlength{\abovecaptionskip}{0.2cm}
\includegraphics[width=7.75cm,height=7.45cm]{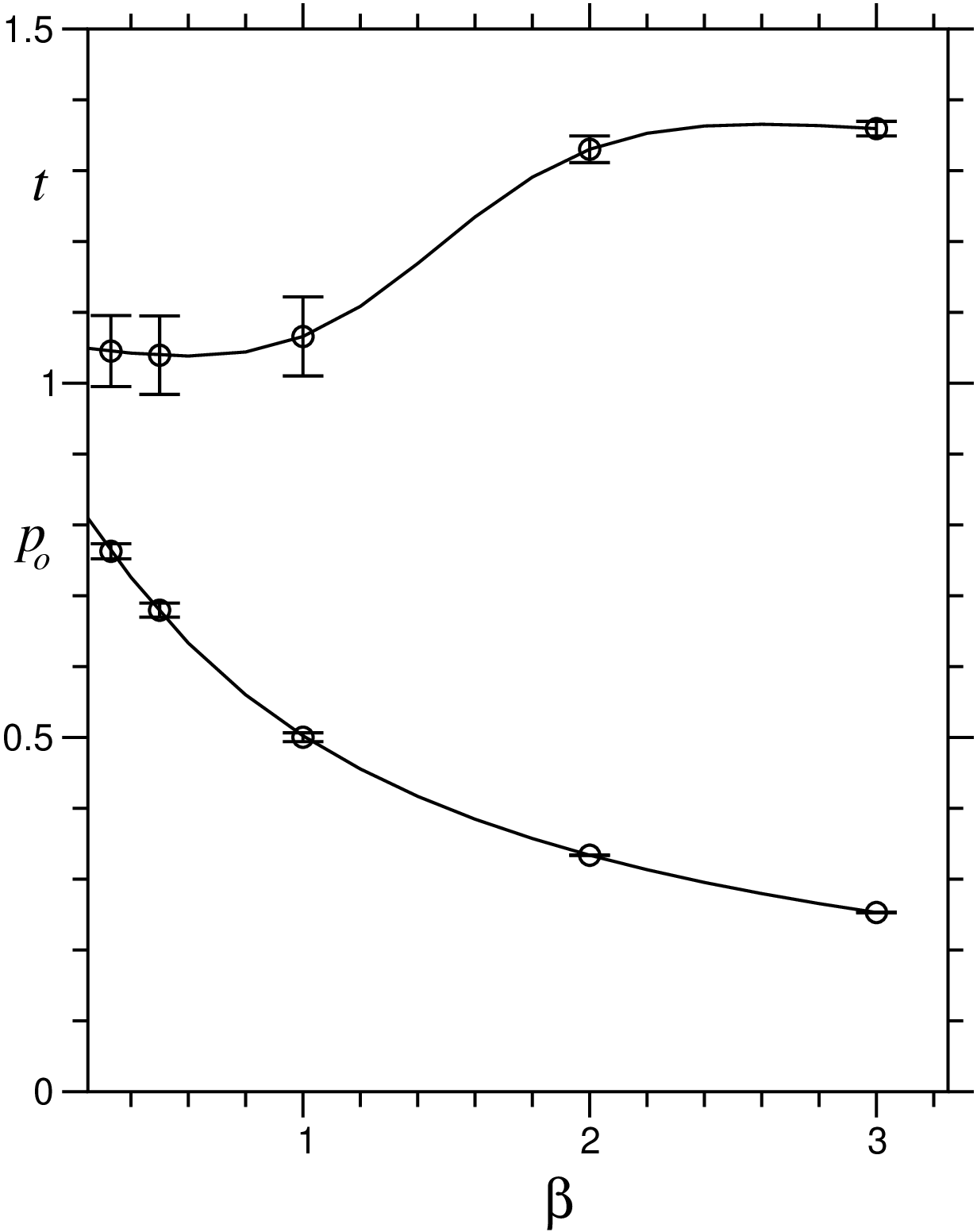}
\caption{\label{fig:epsart405} Values of $t(\beta)$ and $p_0(\beta)$
with $\beta$, for anisotropic removal of conductance. Elements in
the vertical and horizontal directions are removed with
probabilities $p$ and $\beta\,p$.}
\end{figure}

We now validate our results using finite size scaling to independently
estimate $t(0)$. According to the finite size scaling ansatz,
for the correct $t(0)$, the relationship between the rescaled
variables $\overline{\tau}=M^{t(0)/\nu}\times\tau$ and
$\zeta=M^{-1/\nu}(p_{0}-p)$ is independent of the system
size $M$\;\cite{newmanb}. Although the finite size scaling ansatz
need hold only for $p\rightarrow\,p_{0}$ (and
$0<M^{-1/\nu}<1$), we find that the data collapses to a scaling
function $\overline{\tau}(\zeta)$ over the entire range
$p\in[0,p_{0}]$; see Figure~\ref{fig:epsart3} (left side). We then determine the best
value for $t(0)$: for any chosen value of $t(0)$, we
approximate the scaling function $\overline{\tau}(\zeta)$ by a
rational function $\overline{f}(x)/\overline{g}(x)$ (where
$\overline{f}(x)$ and $\overline{g}(x)$ are polynomials of order
$3$ and $2$ respectively) and estimate the deviation of the data
$(\zeta_{k},\overline{\tau}_{k})$ from the scaling function by
$\chi^{2}=\sum_{k}(\overline{\tau}_{k}-\overline{\tau}(\zeta_{k}))^{2}$.
Here, the sum is over all available networks. Figure~\ref{fig:epsart3} (right side) shows
the $\chi^{2}$ as a function of $t(0)$; the best estimate,
which we assume minimizes $\chi^{2}$, is
$t_{FSS}=1.07\pm\,0.10$, where the error estimate corresponds
to doubling the $\chi^{2}$ value. This estimate agrees with that value
we obtained from Equation (1). We tested all remaining estimates
for parameters using Equation (1) and the finite size scaling
ansatz.

\begin{figure}[b]
\centering \vspace*{0.2cm} \setlength{\abovecaptionskip}{0.2cm}
\includegraphics[width=7.75cm,height=7.45cm]{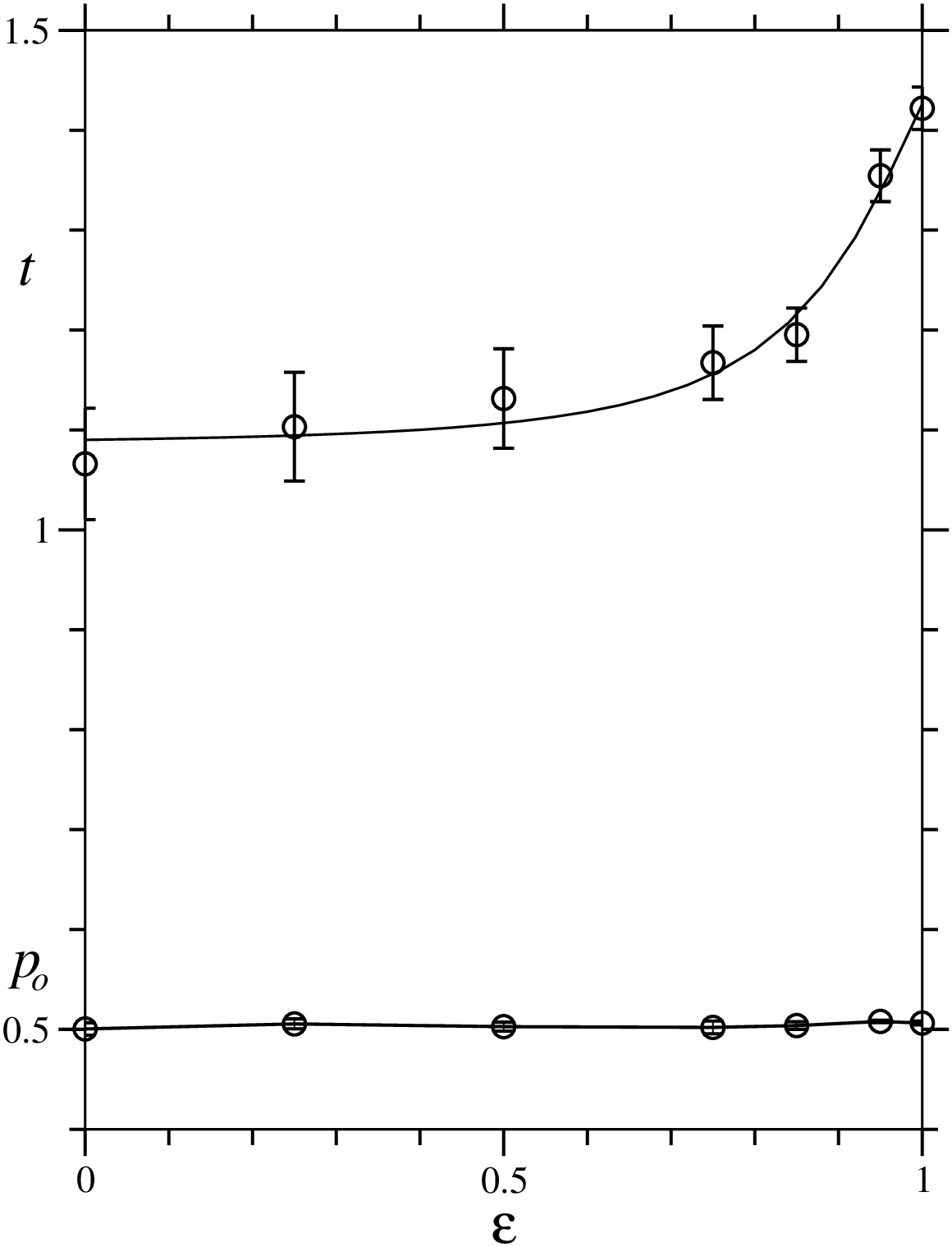}
\caption{\label{fig:epsart5} Behavior of $t(\varepsilon)$ and
$p_0(\varepsilon)$ with $\varepsilon$, when conductances of the initial
network are chosen randomly within $(1-\varepsilon,1+\varepsilon)$.
Bonds removal is isotropic ($p_0$ remains at 0.5).}
\end{figure}

Next we consider networks from which we removed elements
anisotropically. We begin with a square network of unit
conductance and remove elements in the horizontal and vertical
directions with probabilities $p_{h}=\beta\,p$ and
$p_{v}=p$. Analysis of such networks for $\beta=2.0$ using
Equation (1) gives
$\,p_{0}=0.3383\pm\,0003,\,t=1.3303\pm\,0.0189$. Similarly,
for $\beta=1/2\,$,
$p_{0}=0.6798\pm\,0.0099,\,t=1.0396\pm\,0.0553$ (See Figure~\ref{fig:epsart405}). Finite size scaling for the two cases estimates give
$t_{FSS}=1.3\pm\,0.10$ and $t_{FSS}=1.05\pm\,0.10$
respectively, in good agreement with Equation (1). The estimates
for $p_{0}$ for multiple $\beta$'s, shown in Figure~\ref{fig:epsart405}, agree with
theoretical results for anisotropic networks\;\cite{redneru}. The
value of the critical exponent changes little over the range from
$\beta\,>\,1.0\,$ to $\beta\,<\,2.0$.

Now, we consider the slow breaking  network process which consist of two subsequent independent random processes, i.e., after a random remotion of conductors; it is imposed 
to the network remaining elements a conductivity value choosing at random in a range $(1-\varepsilon,1+\varepsilon)$; where $\varepsilon\in\,(0,1)$.  
Note that,  the isotropic case is retrieved for $\varepsilon\,=\,0$. We find that the critical index depends on
the value of $\varepsilon$. We conduct our analysis for
$\varepsilon=0.00,0.25,\,0.50,\,0.75,\,0.85,\,0.95$ and $1.0$, and at each
$\varepsilon$, for network sizes considered earlier. Although the
value of the critical point $p_{0}$ is independent of $\varepsilon$
(Figure~\ref{fig:epsart5}), and the critical exponent increases with $\varepsilon$ (for $\varepsilon\,>\,0.5\,$)
(Figure~\ref{fig:epsart5}), consistent with results for finite size scaling.

\begin{figure}[b]
\centering \vspace*{0.2cm} \setlength{\abovecaptionskip}{0.2cm}
\includegraphics[width=7.75cm,height=7.45cm]{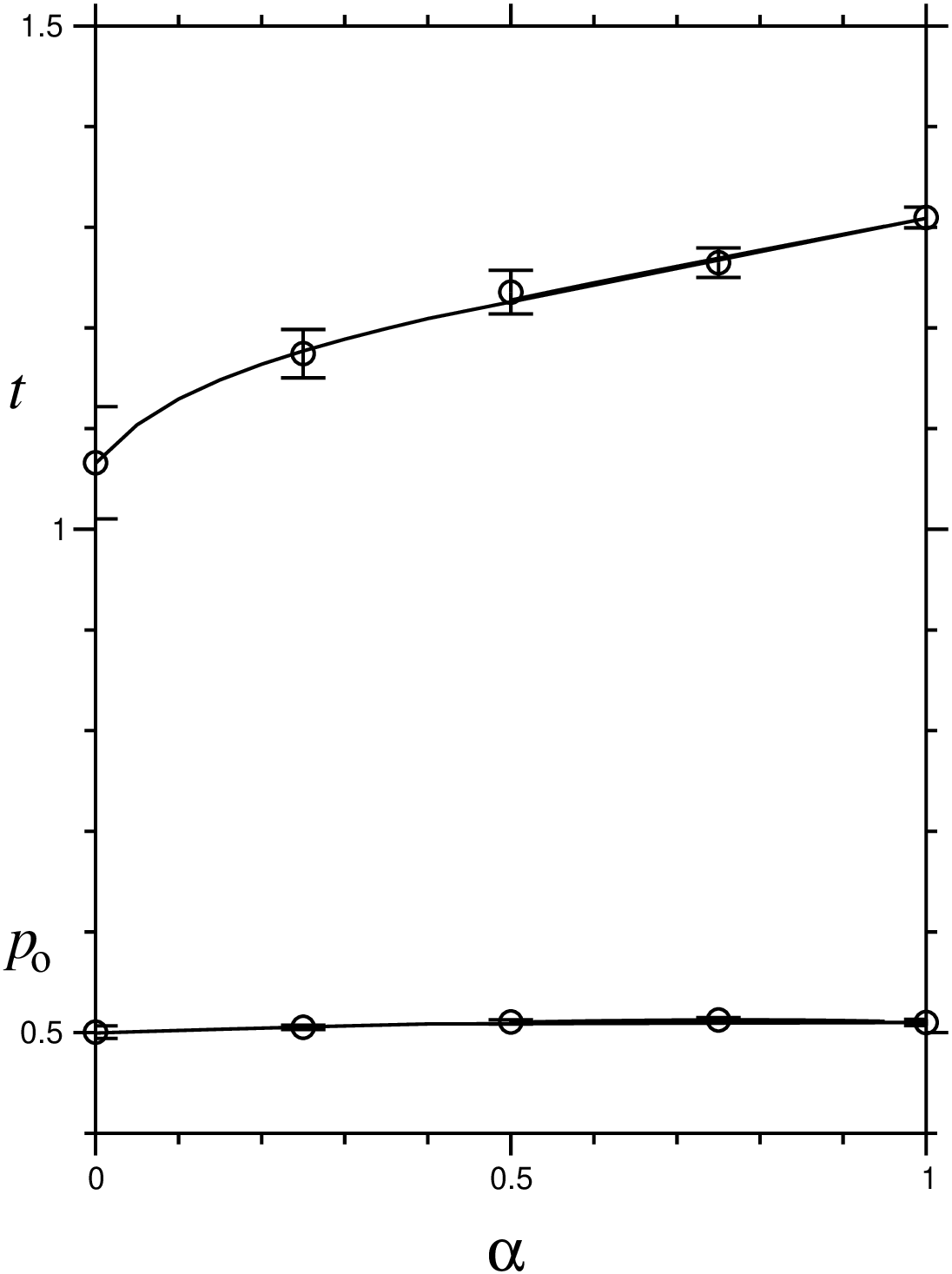}
\caption{\label{fig:epsart604} Variations in parameters $t(\alpha)$
and $p_0(\alpha)$. When elements are removed from the network with
probability $p$, the conductance of those remaining are reduced
simultaneously by a factor $(1-\alpha\,p)$. The critical fraction
removed elements remain 0.5.}
\end{figure}

Next, we consider networks whose remaining electrical
elements randomly degrade as we remove conductances. This problem relate
to degradation of porous bone with aging\;\cite{mosekilde}.
Specifically, we consider networks whose conductance and
breakdown decrease by a factor $(1-\alpha\,p)$, where as before,
$p$ is the probability for an element to be removed from the
network. Once again, as earlier, we analyzed network sizes
ranging from $M=20$ to $M=160$ (Figure~\ref{fig:epsart604}). For $\alpha=1.0$ we find
that $p_{0}=0.509\pm\,0.005$, $t=1.287\pm\,0.034,\,
a_{1}=-0.123\pm\,0.006\,$ and $\,a_{2}=0.067\pm\,0.003$. The estimated
critical exponent using finite size scaling is $t_{FSS}=1.29\pm\,0.10$.

\begin{figure}[t]
\centering\hspace*{-.5cm} \vspace*{0.2cm} \setlength{\abovecaptionskip}{0.2cm}
\includegraphics[width=4.375cm,height=5.75cm]{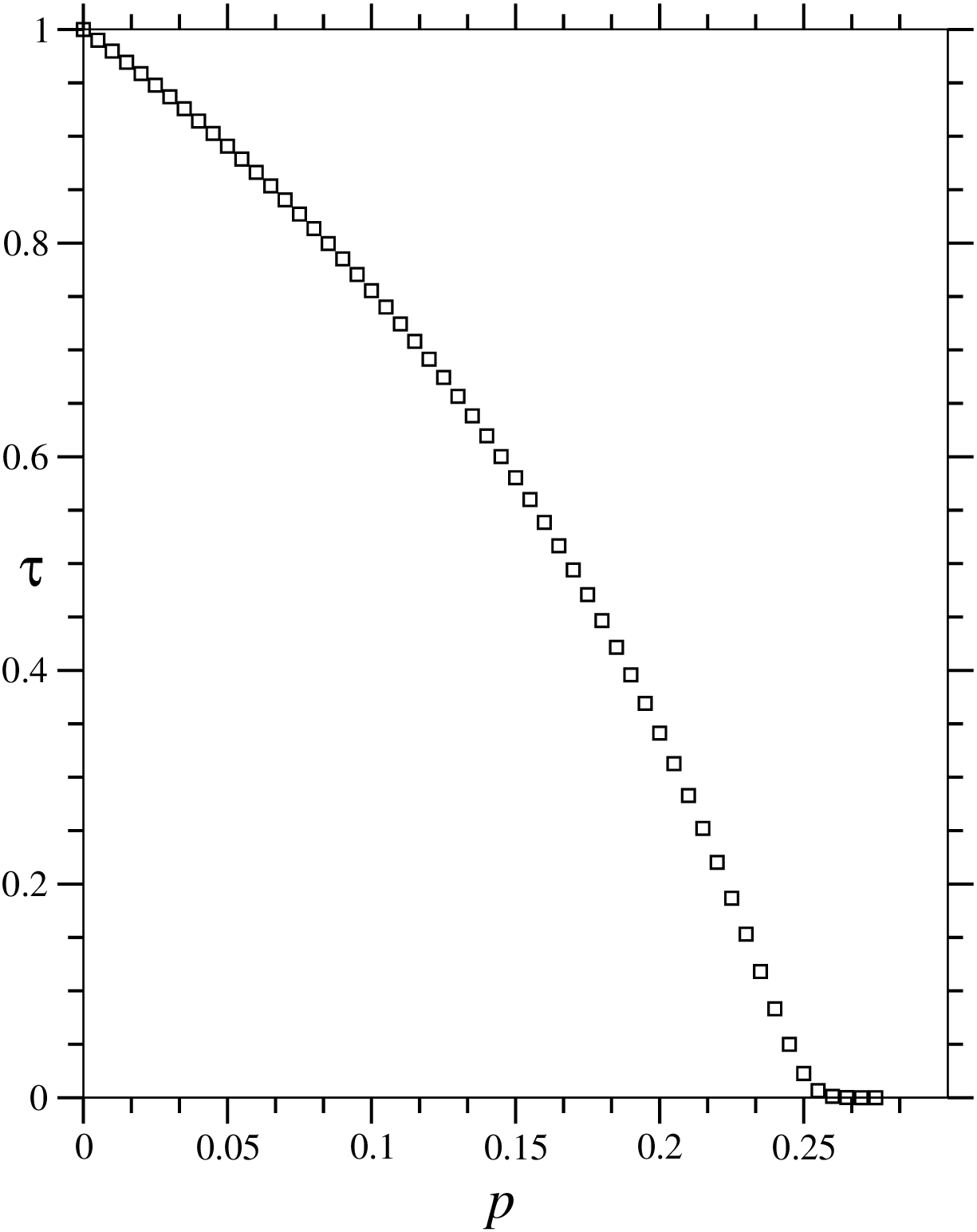}
\includegraphics[width=4.375cm,height=5.75cm]{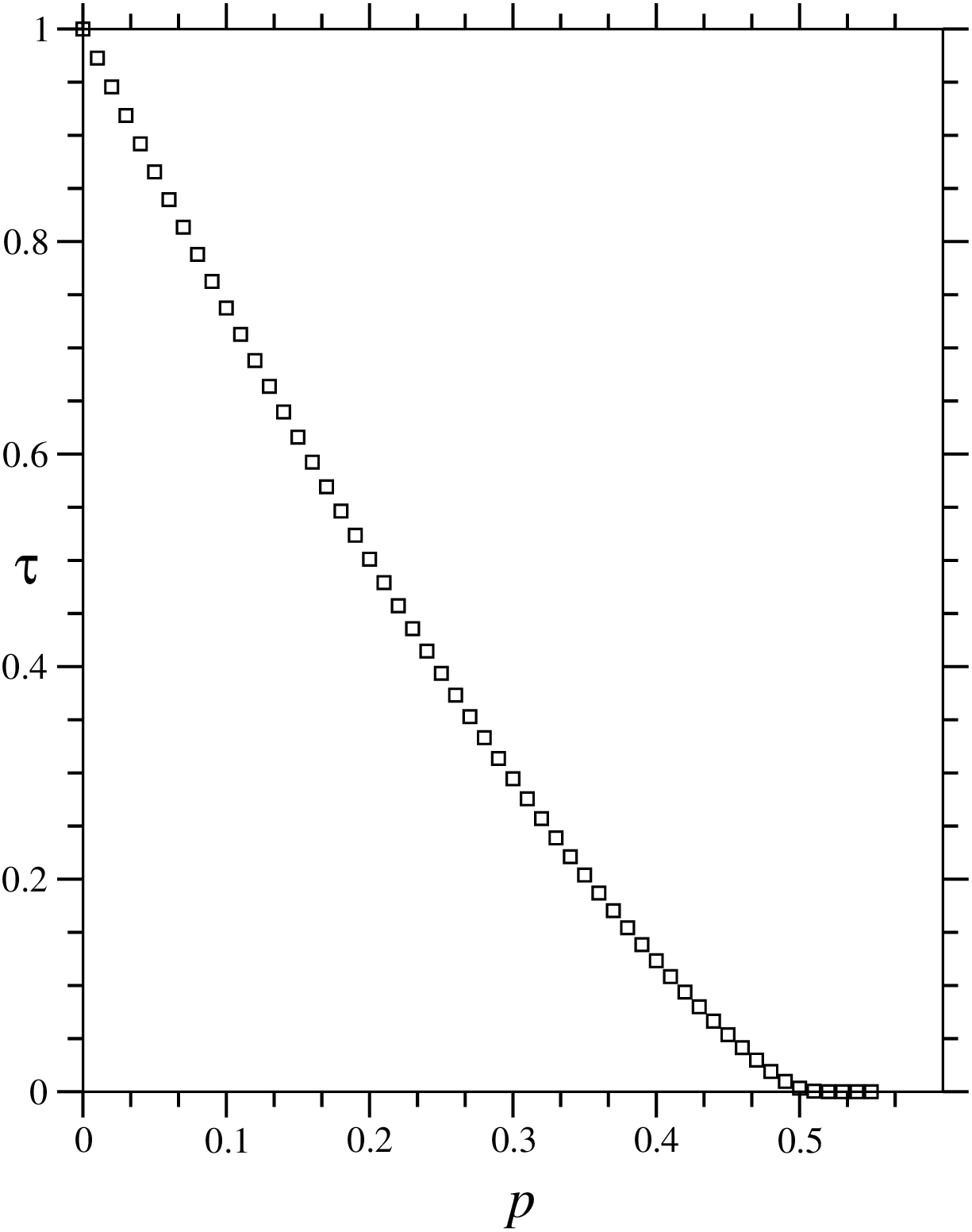}
\caption{\label{fig:epsart7}The deviation from linearity of the strength reduction due to the removal of a fraction of conductance in a $160^{2}$ network.
The right side shows the anisotropic removal of conductance with probabilities $p$ and $3.0\,p$ in the vertical and horizontal direction.The left side corresponds the case where the elements are randomly removed with probability $p$ meanwhile the ones remaining are simultaneously reduced by a factor $(1-0.75\,p)$.}
\end{figure}

Let us now stress a point on the essential of nonlinear and irreversible properties
of the breakdown process.  In Figure~\ref{fig:epsart7} are shown the strength reduction behavior originated by
two different types of damage modalities. As can be seen, both deviates from linearity above
the percolation threshold in agreement with\;\cite{chubThorp}.
The right side of the figure shows the strength reduction behavior for the anisotropic case for $\beta=3.0$. 
Left side of the figure considers an ensemble of conductor  which  are
randomly removed from the network while their remaining 
elements are diminished at random by a factor $(1-\,0.75\,p)$.  For the slow breaking process, the strength reduction  shows a similar behavior. 

\section{Conclusions}
We have computed critical indices for several classes of square
networks of conductances  using finite size scaling and Equation
(1).  For the isotropic case, $t=1.066\,\pm\,0.056$ in agreement with
the value $t=1.1$ reported by Kirkpatrick\;\cite{kirkpatrickc2},
Straley\;\cite{stra1,stra2}, and Stinchcombe and Watson\;\cite{stiAwat}.
However, it is different from the values (close to 1.3) obtained
from real space renormalization group
methods\;\cite{bern,reyAkle,frankc}. (This discrepancy
has already been discussed by Straley\;\cite{stra1}.)
For anisotropic networks, we compare to an analysis of
experimental and computational results by Han, Lee and Lee\;
\cite{hanu} and by Smith and Lobb\;\cite{smithnu}. Both groups
found that $t=1.3$ when $p=0.33$. Figure~\ref{fig:epsart405} shows that for
$p=0.33$, the parameter $\beta$ is $1.94$ and $t=1.31$. Further, the values
of $p_0(\beta)$ for anisotropic removal of conductances (Figure~\ref{fig:epsart405}) is consistent with theoretical results of Redner and Stanley
\;\cite{redneru}.

Here, we have presented substantial evidence that critical indices depend on the type of initial network and
damage modalities of conductances and the network. 
Indeed the critical exponent might be a convoluted exponent because of two independent random processes are affecting the strength reduction of networks.
Results from previous studies have been shown to be isolated examples of our
more general analysis of this problem.  It would be of interest to develop a renormalization group based
analysis to describe these more general damage processes. We have, thus
far, not been successful in identifying such a theory.
\begin{acknowledgments}
The authors would like to thank Dr. K.E. Bassler for useful discussions and to Dr. R.E. Lagos for carefully reading this manuscript.
This work was partially funded by the National Science Foundation, the Institute of Space Science Operations and the
ICSC-World Laboratory.
\end{acknowledgments}

\bibliography{JSEO_V2008}

\end{document}